\documentclass{article}

\usepackage{graphicx}
\usepackage[utf8]{inputenc}
\usepackage[vlined,ruled, procnumbered]{algorithm2e}
\usepackage{algcompatible}
\usepackage{newfloat}
\usepackage{graphicx}% Include figure files
\usepackage{epstopdf}
\usepackage{dcolumn}% Align table columns on decimal point
\usepackage{bm}% bold math
\usepackage{mathtools}
\usepackage[]{subfigure}
\usepackage{cite}

\usepackage{color}

\begin{document}

\title{Time Centrality in Dynamic Complex Networks}% Force line breaks with \\

\author{Eduardo C. Costa; Alex B. Vieira\\
Computer Science Dept\\ Universidade Federal de Juiz de Fora (UFJF)\\36036-330, Juiz de Fora, MG, Brazil\\
\texttt{eduardo.chinelate@gmail.com; alex.borges@ufjf.edu.br}\\
\\
Klaus Wehmuth; Artur Ziviani\\
Computer Science Dept\\ National Laboratory for Scientific Computing (LNCC)\\25651-075, Petr\'{o}polis, RJ, Brazil\\
\texttt{klaus@lncc.br; ziviani@lncc.br}\\
\\
Ana Paula Couto da Silva\\
Computer Science Dept\\ Universidade Federal de Minas Gerais (UFMG)\\31270-010, Belo Horizonte, MG, Brazil\\
\texttt{ana.coutosilva@dcc.ufmg.br}
}

\date{}

\maketitle

\begin{abstract}
%\art{Ainda vou rever}
There is an ever-increasing interest in investigating dynamics in time-varying graphs (TVGs). 
Nevertheless, so far, the notion of centrality in TVG scenarios usually refers to metrics that assess the relative importance of nodes along the temporal evolution of the dynamic complex network. For some TVG scenarios, however, more important than identifying the central nodes under a given \emph{node centrality} definition is identifying the \emph{key time instants} for taking certain actions. In this paper, we thus introduce and investigate the notion of \emph{time centrality} in TVGs. Analogously to node centrality, time centrality evaluates the relative importance of time instants in dynamic complex networks. In this context, we present two time centrality metrics related to diffusion processes. We evaluate the two defined metrics using both a real-world dataset representing an in-person contact dynamic network and a synthetically generated randomized TVG. We validate the concept of time centrality showing that diffusion starting at the best classified time instants (i.e. the most central ones), according to our metrics, can perform a faster and more efficient diffusion process. 
\end{abstract}

\section{Introduction}

%The concept of centrality in graph theory and network analysis usually refers to metrics that assess the relative importance of nodes within a graph~(network). Different ways of measuring node centrality have been proposed for decades, addressing node centrality from a number of different perspectives and applications~\cite{freeman1978,freeman1977,sabidussi1966}. Examples include using node centrality to evaluate network robustness to fragmentation or to identify the most important nodes for efficient information spreading \cite{xuan2003, khelil2002}. 

We witness an ever-increasing interest in investigating the dynamics of complex networks (i.e., changes in nodes or edges over time) representing them as time-varying graphs~(TVGs) \cite{kostakos2009,holme2012,casteigts2013}. In this context, a number of recent efforts investigate new centrality definitions to capture the relative node importance in TVGs~\cite{braha2009,tang2010,lerman2010,udin2011,kim2012,nicosia2013,tang2013}. 
%For instance, one can target the identification of the most strategic nodes for information spreading in TVGs.
For some TVG scenarios, however, more important than identifying the central nodes under a given definition, is identifying the \emph{key time instants} for taking certain actions. Deciding \emph{when}, and not only where from, to start a diffusion process can be of upmost importance for a more effective outcome. 
For example, Spasojevic~et~al.~\cite{spasojevic2015} has recently shown evidence that recommending best times for a user to post on social networks improves the probability of audience responses.
%For example, it is interesting to determine the best time instant at which a vaccine should be applied in a population to achieve better responses to an epidemic spreading.

This paper introduces and investigates the notion of \emph{time centrality} in dynamic complex networks. Analogously to node centrality, time centrality assesses the relative importance of a given time instant within a TVG, which can be defined in different ways.
%As with node centrality, time centrality can be defined in different ways. 
%For instance, in an information diffusion process, it could be more strategic to identify the best moment to start spreading information for a faster diffusion than identifying the best node to start the diffusion.
 In this context, we present two metrics, namely \emph{cover time} and \emph{time-constrained coverage}, to assess time centrality in TVGs from different perspectives. In doing so, we also show the relevance and relative importance (i.e. centrality) that some time instants may reach in comparison with other (less important) time instants. Both metrics regard different perspectives related to diffusion processes in complex dynamic networks. 
 
We evaluate both time centrality metrics we define relying on a real-world dataset of an in-person contact network in a hospital environment as well as on a synthetically generated randomized TVG~(details in Section~\ref{results}). Our results show that start spreading information at the top ranked (i.e., the most central) time instants, according to our metrics, can make diffusion processes in TVGs faster and more efficient. 
For example, compared with a random choice, a diffusion process based on time centrality metrics in the adopted real-world dataset can reach 10\% of the studied TVG about 3 times faster in the median case. Likewise, a diffusion process starting at the most central time instants can cover more nodes in the analyzed TVGs with a limited number of allowed time steps for spreading when compared with random chosen time instants.

The paper is organized as follows. Section~\ref{relWor} discusses related work on temporal centrality in complex networks. In Section~\ref{UMRTVG}, we present the model we use to represent TVGs. 
Section~\ref{TC} introduces the notion of time centrality and formalize two time centrality metrics related to diffusion processes. In Section~\ref{results}, we discuss results on analyzing time centrality in a real-world dataset representing a in-person contact network and a randomized TVG. Finally, we summarize results and present the outlook of future work based on time centrality in Section~\ref{conclusion}.

\section{Related Work}
\label{relWor}

A large number of works on dynamic complex networks concerns both data structures to model TVGs and different definitions of node centrality in that context. 
The first works that attempted to represent TVGs were influenced by traditional graph modeling. For instance, a number of works model TVGs through a sequence 
of static subgraphs that represents the network dynamics in a discrete way as time passes~\cite{xuan2003,ferreira2004,holme2005,holme2012,tang2009,tang2010,tang2010b}.

Recent works tend to represent TVGs in different manners. For example, Casteigts \textit{et al.}~\cite{casteigts2013}
use a presence function defined over continuous time intervals. The model proposed by Kostakos~\cite{kostakos2009} is based on the idea that a class of links represent instantaneous interations between distinct nodes while another class represents a waiting state of a given node. Kim and Anderson~\cite{kim2012} suggest the use of edges connecting nodes at different time instants~(temporal and mixed dynamic edges).
For any of these TVG models, recent proposals investigate adaptations of the node centrality notion from traditional graphs to the time-varying context, leading to different notions of temporal node centrality targeted at particular applications~\cite{braha2009,tang2010,lerman2010,udin2011,kim2012,nicosia2013,tang2013}.

As far as we know, this paper is the first to propose the notion of time centrality in dynamic complex networks. 
%In fact, as discussed, related works are more focused on assessing node centrality in complex networks than determining the most important time instants in a TVG. 
In contrast to previous work, this paper intends to assess the most important time instants, thus exploring the notion of time centrality.

\section{Modeling Time-Varying Graphs}
\label{UMRTVG}

We model TVGs as a particular case of a MultiAspect Graph~(MAG)~\cite{wehmuth2014a,wehmuth2015a} in which the vertices and time instants are the key features (i.e., aspects) to be represented by the model. 
A MAG is a structure capable of representing multilayer and time-varying networks while also having the property of being isomorphic to a directed graph.
The MAG structural form resembles the multilayer structure recently presented by~\cite{kivela2014}, since in both cases the proposed structure has a construction similar to an even uniform hypergraph associated with an adjacency concept similar to the one of simple directed graphs. 

Formally, a MAG can be defined as an object $H=(A,E)$, where $E$ is a set of edges and $A$ is a finite list of sets, each of which is called an aspect. In our case, for modeling a TVG, we have two aspects, namely vertices and time instants, i.e. $|A|=2$. For the sake of simplicity, this 2-aspect MAG can be regarded as representing
a TVG as an object $H = (V, E, T)$, where $V$ is the set of nodes, $T$ is the set of time instants, and $E \subseteq V \times T \times V \times T$ is the set of edges. 
%The sets $V$ and $T$ are the two aspects in the aspect list~$A$.
As a matter of notation, we denote $V(H)$ as the set of all nodes in $H$, $E(H)$ the set of all edges in $H$, and $T(H)$ the set of all time instants in $H$. %We also define $n(H) = |V(H)|$ the number of nodes in $H$, $m(H) = |E(H)|$ the number of edges in $H$, and $\tau(H) = |T(H)|$ the number of time instants in which $H$ is defined.

An edge $e \in E(H)$ is defined as an ordered quadruple $e = (u, t_a, v, t_b)$, where $u, v \in V(H)$ are the origin and destination nodes, 
%($u$ possibly equal to $v$), 
respectively, while $t_a, t_b \in T(H)$ are the origin and destination time instants, respectively.
%($t_a$ possibly equal to $t_b$). 
Therefore, $e = (u, t_a, v, t_b)$ should be understood as a directed edge from node $u$ at time $t_a$ to node $v$ at time $t_b$. If one needs to represent an undirected edge in the TVG, both $(u, t_a, v, t_b)$ and $(v, t_b, u, t_a)$ should be in $E(H)$. 

We also define a \emph{temporal node} as an ordered pair $(u, t_a)$, where $u \in V(H)$ and $t_a \in T(H)$. 
The set $VT(H)$ of all temporal nodes in a TVG $H$ is given by the cartesian product of the set of nodes and the set of time instants, i.e. $VT(H) = V(H) \times T(H)$. 
As a matter of notation, a temporal node is represented by the ordered pair that defines it, e.g. $(u, t_a)$. 

The usage of the object $H=(V,E,T)$ to represent a TVG is formally introduced in~\cite{wehmuth2015b}. Therein, the representation of the TVG based on temporal nodes is proven to be isomorphic to a directed static graph. This is an important theoretical result since this allows the use of the isomorphic directed graph as a tool to analyze both the properties of a TVG and the behavior of dynamic processes over a TVG, as done is this work. This model is also shown to unify the representation of several previous (classes of) models for TVGs of the recent literature, which in general are unable to represent each other~\cite{wehmuth2015b}.

Importantly, the adopted TVG model is a particular case of a MAG~\cite{wehmuth2014a,wehmuth2015a}. Therefore, adding new aspects, such as different network layers, is straightforward. This opens perspectives of extending the analysis of the time centrality concept we introduce in Section~\ref{TC} to more complex dynamic networked systems, such as time-varying multilayer networks~\cite{wehmuth2014a,kivela2014,boccaletti2014}. We intend to further explore such perspectives in our future work. 

In the context of this paper, a time instant is equivalent to a snapshot in the TVG. Of course, defining the snapshot faces similar problems as any discretization process. For real datasets, snapshots covering large time intervals may not adequately capture the topology dynamics, whereas snapshots covering too short time intervals may generate excessive representation data without relevant information. To avoid these issues, in this paper, we chose to have snapshots corresponding to the event granularity in the case of the realworld dataset, i.e. 30s. Similarly, in the case of the randomized TVG, the granularity of the  snapshots  is equivalent to the  changes in the topology in an atemporal way since we consider a new random graph in each snapshot. These TVGs used for the evaluation are explained in further detail in Section~

\section{Time Centrality}
\label{TC}

%Traditionally, there is interest in defining nodes importance within a system. Even in TVGs, node centrality is challenging as one can consider a node importance under different point of views.  Furthermore, TVGs allow us to observe and evaluate distinct features of networks. For instance, one can try to find the most import moment during a system existence.

We here introduce the notion of \textit{time centrality}. 
%Analogously to node centrality, our goal is to assess the relative importance of distinct time instants of a TVG.
As with node centrality, different definitions of time centrality can reflect different notions of importance for time instants targeting distinct applications.
In other words, as node centrality creates a node ranking adopting an arbitrary importance criterion, time centrality creates a ranking of time instants using some importance criterion to measure the relative importance a time instant may have when compared with others.
%through a particular centrality metric. 
%For example, depending on the application, one can define time centrality considering the number of connected nodes or the number of edges a time instant has; or simply take into account if a diffusion process would be more effective if it starts at that time instant.
 
We present two time centrality metrics that can assess diffusion in TVGs from different perspectives. To define such metrics, we consider a generic TVG $H=(V,E,T)$~(see Section~\ref{UMRTVG}). Let $N=|T(H)|$ be the size of set $T(H)$.  Each time instant $t_i \in T(H)$, where $0 \leq i < N$, composes the set of sequential time instants described by $T(H) = \{t_0, t_1, \dots, t_{N-1}\}$. For the sake of clarity, the illustrative figures in this section show sequences of time instants as disconnected \emph{snapshots}, although in the adopted model temporal edges connect nodes in different time instants, easing the temporal analysis. 

\subsection{Cover Time (CT)}
\label{CT}

The \emph{Cover Time (CT)} metric from a time centrality perspective assesses the minimum number of time instants~(steps) a diffusion takes to reach a given fraction of the nodes. More formally, the cover time $CT(t_i,\tau)$ returns the average amount of time instants~(steps) among all nodes for a diffusion starting at time $t_i$ to reach a given fraction $\tau$ of the complex dynamic network, i.e.

\begin{equation}
CT(t_i, \tau) = 
\begin{dcases}
	\frac{1}{|V(H)|} \sum_{u \in V(H)} d_t(t_i,u,\tau), & \text{if $\tau$ reached,}\\
	\infty, 								& \text{otherwise.}
\end{dcases}
\end{equation}

\noindent
where $d_t(t_i,u,\tau)$ is the number of time instants (steps) for a diffusion starting on node $u$ at time instant $t_i$ to reach a fraction $\tau$ of nodes. If the fraction $\tau$ of nodes is not reached for a diffusion process starting on any node at $t_i$, then $CT(t_i, \tau)$ is considered $\infty$. The results presented in Section~\ref{results} consider only the finite results of cover time.

We consider the diffusion process to follow a Breadth-First Search~(BFS), i.e. a generic node $(u, t_i)$ starts diffusion by sending information to all of its adjacent nodes. Then, the adjacent temporal nodes relay information for their own adjacent nodes in the next time instant, and so on. Information is distributed until the fraction $\tau$ of nodes is reached and the time needed for diffusion is stored. This evaluation is repeated for every node $(u, t_i) \in VT(H)$. The average diffusion time is the cover time centrality of the time instant $t_i$.

\begin{figure}[ht]
\begin{center}
\includegraphics[width=0.7\columnwidth]{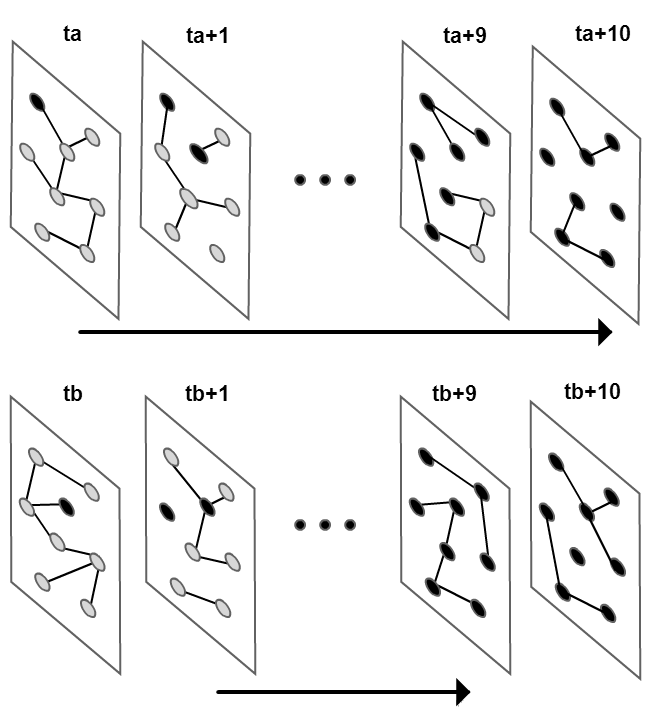}
\caption{Illustrative cover time on TVGs.}
\label{diffusion}
\end{center}
\end{figure}

As an example, the top part of Figure~\ref{diffusion} shows an illustrative diffusion process from a single node starting at $t_a$ and finishing at $t_{a+10}$, whereas the bottom part of Figure~\ref{diffusion} presents another diffusion process starting at $t_b$ and finishing at $t_{b+9}$. Intuitively, the time instant $t_b$ is more important (central) in regard of its cover time, since the diffusion starting at it is faster. Recall that the CT metric actually considers an average of the diffusion processes starting at each node.

\subsection{Time-Constrained Coverage (TCC)}

\begin{figure}[ht]
\begin{center}
\includegraphics[width=0.9\columnwidth]{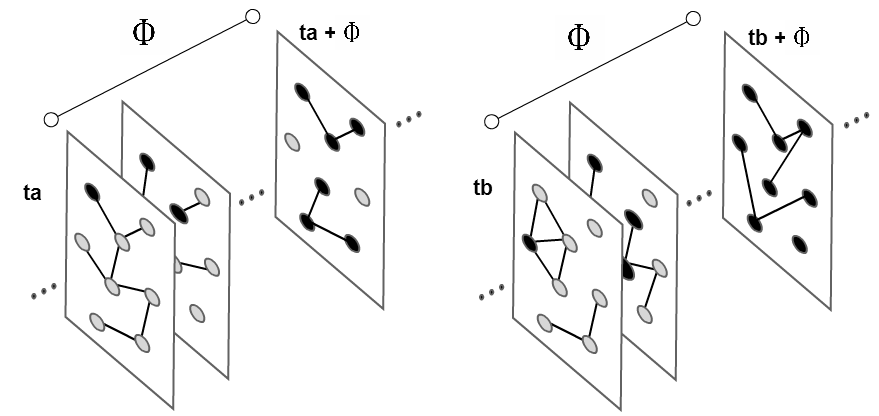}
\caption{Illustrative time-constrained coverage on TVGs.}
\label{coverage}
\end{center}
\end{figure}

%For each time instant $t_i$, the \emph{Time-Constrained Coverage (TCC)} returns the average percentage of TVG nodes that is reached by the information diffusion process, given a limited number of diffusion steps $\Phi$. More precisely, 
The \emph{Time-Constrained Coverage (TCC)} metric from a time centrality perspective evaluates the TVG coverage achieved by a diffusion process after a limited number of time steps.
Here we assume the same diffusion process defined for the cover time~(Section~\ref{CT}). More precisely, for a given time instant $t_i$, $TCC(t_i, \Phi)$ returns the average fraction of the TVG nodes reached by the diffusion process in $\Phi$ of time steps. More formally, 

\begin{equation}
TCC(t_i, \Phi) = \frac{1}{|V(H)|^2} \sum_{u \in V(H)} d_c(t_i,u,\Phi),
\end{equation}

\noindent
where $d_c(t_i,u,\Phi)$ is the number of nodes reached from node $u$ after $\Phi$ time steps when the diffusion starts at time $t_i$.
For example, Figure~\ref{coverage} presents two distinct illustrative diffusion processes occurring in $\Phi$ time steps from a single node. At the left part, the process starting at $t_a$ reaches less nodes than the process starting at $t_b$ at the right part after $\Phi$ time steps. The latter diffusion process actually reaches all nodes in the TVG at the $t_{b+\Phi}$ time instant.
Therefore, the time instant $t_b$ is more important (central) in regard of its time-constrained coverage, since the diffusion starting at it is more effective in reaching a larger number of nodes after $\Phi$ time steps.
Again, recall that the TCC metric considers an average of the diffusion processes starting at each node.

\section{Time Centrality Results}
\label{results}

In this section, we first present the TVGs we use to evaluate the two time centrality metrics. We next analyze the results for the cover time metric. Then we analyze the results for the time-constrained coverage metric. Finally, we further discuss on the achieved outcome of the time centrality metrics as well as on some limitations of the study. 

\subsection{Real-world dataset and randomized TVG}
\label{tvgs-evaluation}

We evaluate the time centrality metrics we propose using a TVG dataset collected in the context of the MOSAR project~\cite{lucet2012}.
MOSAR (Mastering hOSpital Antimicrobial Resistance and its spread) is a scientific collaboration project that comprises several medical, biochemistry, and computing research institutions. The MOSAR project focuses on antimicrobial-resistant bacteria (AMRB) transmission dynamics in high-risk environments, such as intensive care units and surgical centres. The adopted dataset consists of the records of in-person contacts (from physicians, nurses, staff members, and patients) in a certain medical ward for a period of two weeks~(between 12 am of July 25, 2009 and 12am of August 08, 2009). Each one of the 160 volunteers who participated in the study was equipped with a RFID device that detected the presence of another devices within a small distance~(about one meter). Device identification was unique and was always associated with the same person. Every 30 seconds during the two-weeks period, each device registered the list of all devices (nodes) that were within its coverage area in order to establish the arrangement of the contacts among them (edges). 

We have modeled the MOSAR dataset as a TVG $M = (V, E, T)$, as described in Section~\ref{UMRTVG}, where $V(M)$ is the set of all participants, $E(M)$ is the set of all in-person contacts, and $T(M)$ is the set of sequential time instants at a 30 seconds granularity. Thus, $|V(M)| = 160$ and $|T(M)| = 40320$. Contacts are represented by non-oriented edges, i.e., the edge between any two persons is represented by a directed edge and its reciprocal. 
%Due to the nature of the connections among nodes, the TVG we have generated in this work does not present mixed and self-loop dynamic edges.

We also evaluate the time centrality metrics using a randomized TVG $R = (V, E, T)$, where $V(R)$ is the set of nodes, $E(R)$ is the set of edges, and $T(R)$ is the set of time instants,  represented by a sequence of independent snapshots. Each snapshot is synthetically generated as a random graph based on the Erd\"{o}s-R\'{e}nyi $G = (n, p)$ model, where $n$ is the number of nodes in the graph and $p$ is the probability for edge creation between each pair of nodes. For the randomized TVG, we have kept the same number of nodes of MOSAR TVG, i.e., $|V(R)| = n = 160$, and we have arbitrarily defined the number of time instants $|T(R)| = 800$. Furthermore, we have defined the probability $p = 0.01 \times \frac{\ln 160}{160}$, ensuring that each resulting random graph corresponding to a snapshot presents a similar level of sparsity as compared with the MOSAR TVG (i.e. the value 0.01 ensures this sparsity). This value of probability $p$ also ensures that the generated random graphs are disconnected since it is known that $p > \frac{\ln n}{n}$ is a sharp threshold for the connectedness of $G = (n, p)$ random networks \cite{erdos1960}.

\subsection{Analysis of Cover Time}

\begin{figure}
\centering
\subfigure[Cover time distributions in the MOSAR TVG for varying values of the fraction $\tau$ of nodes to be covered. According to the results, to cover 10\% of the network~($\tau=0.1$), the median cover time is 513 time steps. Similarly, to reach 60\% of the network~($\tau=0.6$), the median cover time is 2994 time steps.]{
   \includegraphics[scale=0.7]{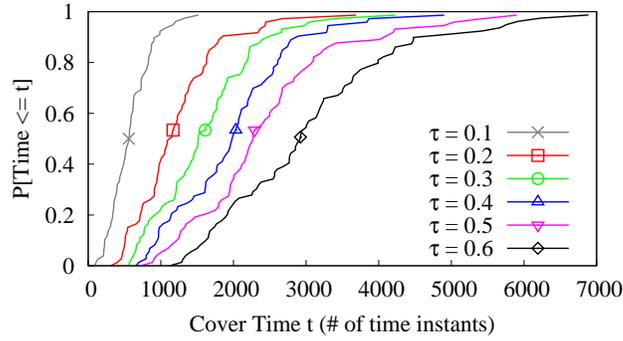}
   \label{cover-time-distribution-mosar}
 }
\subfigure[Cover time distribution in the randomized TVG for varying values of the fraction $\tau$ of nodes to be covered. According to the results, to cover 10\% of network~($\tau=0.1$), the median cover time is 50 time steps. Similarly, to reach 60\% of the network~($\tau=0.6$), the median cover time is 106 time steps.]{
   \includegraphics[scale=0.7]{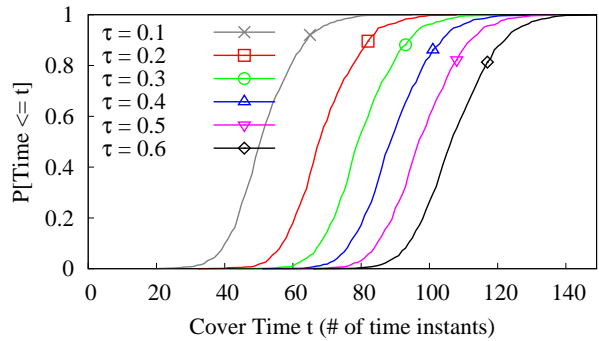}
   \label{cover-time-distribution-random}
 }
\caption{Cover time distributions in MOSAR and randomized TVGs.}
\label{cover-time-distribution}
\end{figure}

We first present results for the cover time metric. As previously discussed, lower cover time values indicate more central time instants that can spread information in a shorter period of time. 
%Moreover, the larger portion of the TVG the spreading intends to achieve, the faster will be the diffusion process.
Figure~\ref{cover-time-distribution-mosar} and~\ref{cover-time-distribution-random} present the cumulative distribution function~(CDF) of the cover time for different values of the fraction $\tau$ of nodes to be covered in the MOSAR TVG and in the randomized TVG, respectively. Six different scenarios in each subfigure are considered with $\tau = \{0.1, 0.2, \dots, 0.6\}$. In Figure~\ref{cover-time-distribution-mosar}, results compare the first 33000 time instants, so there is room for the diffusion to spread through the 40320 time instants of the studied MOSAR TVG. For the reason, in Figure~\ref{cover-time-distribution-random}, results compare the first 660~time instants of the 800~time instants composing the randomized TVG.

In Figure~\ref{cover-time-distribution}, each value in a curve represents the fraction of the studied time instants that achieved the corresponding $\tau$ fraction of nodes in at most the corresponding number of time steps, i.e. the cover time~$t$. In order to improve visualization quality, we add symbols on arbitrary positions of the curves~(the same applies to all remaining figures).
%Each curve represents the probability of random variable Time taking values greater than $t$, where $t$ is the mean cover time that information diffusion processes take to reach $\tau$ \% of network. 
For instance, in Figure~\ref{cover-time-distribution-mosar}, to cover 60\% of the MOSAR TVG~($\tau=0.6$), 20\% of the time instants take less than 2000~time steps. In other words, information diffusion starting at 20\% of the time instants takes less than 16 hours and 40 minutes to cover 60\% of the nodes in the MOSAR TVG only spreading opportunistically through the in-person contact dynamic network. In Figure~\ref{cover-time-distribution-random}, to cover the same fraction of nodes, 20\% of the time instants take less than 100~time steps. 

\begin{figure}
\centering
\subfigure[Cover time in the MOSAR TVG. To reach 10\% of the network~($\tau=0.1$), the median cover time is 206 time steps, if the diffusion starts at the top 10 time instants; whereas the median cover time is 815 time steps, if the diffusion starts at 10 randomly chosen time instants.]{
   \includegraphics[scale=0.7]{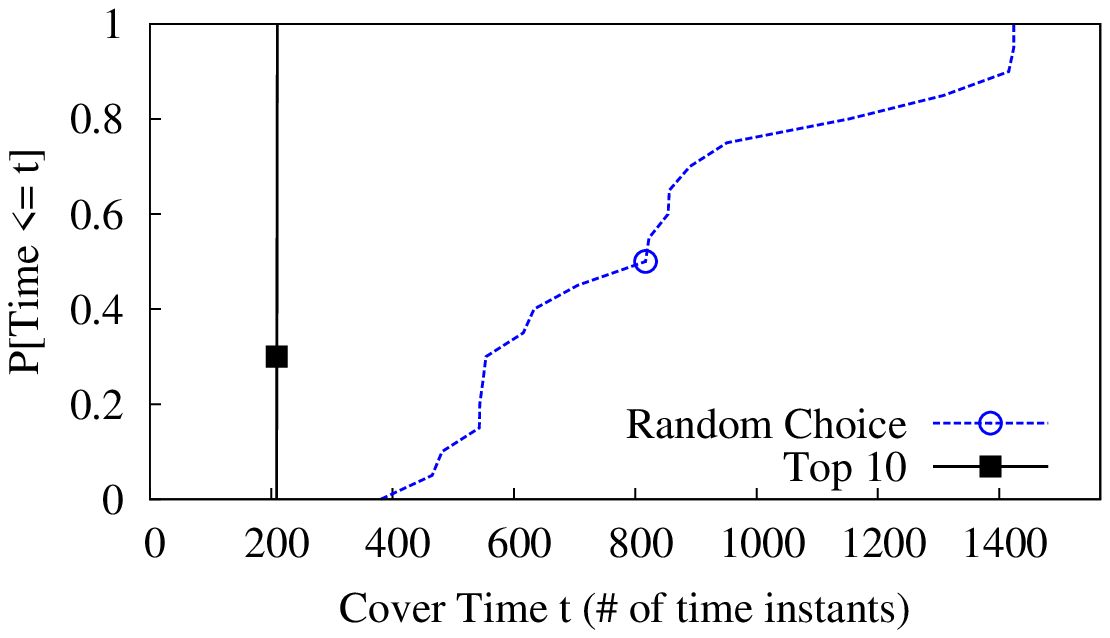}
   \label{cover-time-top10-mosar}
 }
\subfigure[Cover time in the randomized TVG. To reach 10\% of the network~($\tau=0.1$), the median cover time is 27 time steps, if the diffusion starts at the top 10 time instants; whereas the median cover time is 48 time steps, if the diffusion starts at 10 randomly chosen time instants.]{
   \includegraphics[scale=0.7]{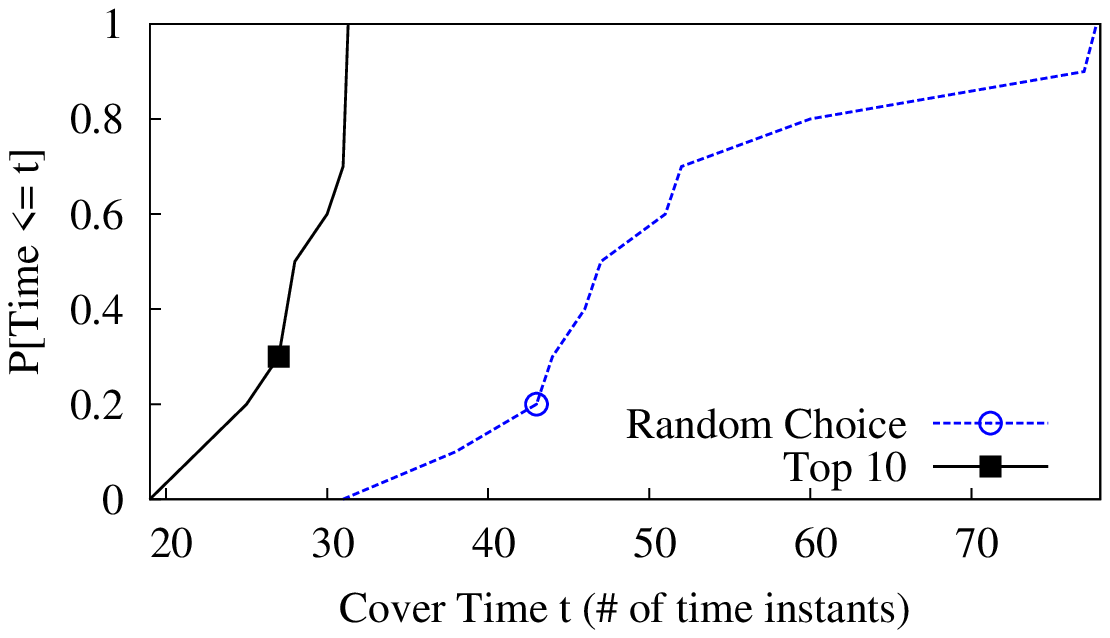}
   \label{cover-time-top10-random}
 }
\caption{Cover time in MOSAR and randomized TVGs: top 10 vs Random Choice.}
\label{cover-time-top10xRandomChoices}
\end{figure}

From Figure~\ref{cover-time-distribution}, for both cases, we clearly note that starting the diffusion at a few central time instants reduces significantly the time of the spreading to reach a certain portion of the nodes in the TVG. Therefore, we also measure the relative improvement of diffusion efficiency achieved by using only the most central time instants. In Figure~\ref{cover-time-top10xRandomChoices}, we compare for both cases, i.e. the MOSAR TVG and the randomized TVG, the performance of the top 10 most central time instants according to the cover time metric against 10 randomly chosen time instants to start spreading a diffusion. Taking the mean result as a reference, to cover 10\% of the nodes~($\tau = 0.1$) in the MOSAR TVG in Figure~\ref{cover-time-top10-mosar}, a diffusion starting at the top 10 most central time instants takes 3 times less time steps than a diffusion starting at the random set of time instants. Similarly, for the randomized TVG in Figure~\ref{cover-time-top10-random}, a diffusion starting at the top 10 most central time instants takes almost half the time steps than a diffusion starting at the random set of time instants to cover 10\% of the nodes~($\tau = 0.1$).

\subsection{Analysis of Time-Constrained Coverage}

\begin{figure}[t]
\centering
\subfigure[Time-constrained coverage distribution in MOSAR TVG. According to the results, for a $\Phi = 500$ time steps diffusion, the median time-constrained coverage is 10\% of network. Similarly, for $\Phi = 2000$ time steps, the median fraction of network reached is 41.5\%.]{
  \includegraphics[scale=0.7]{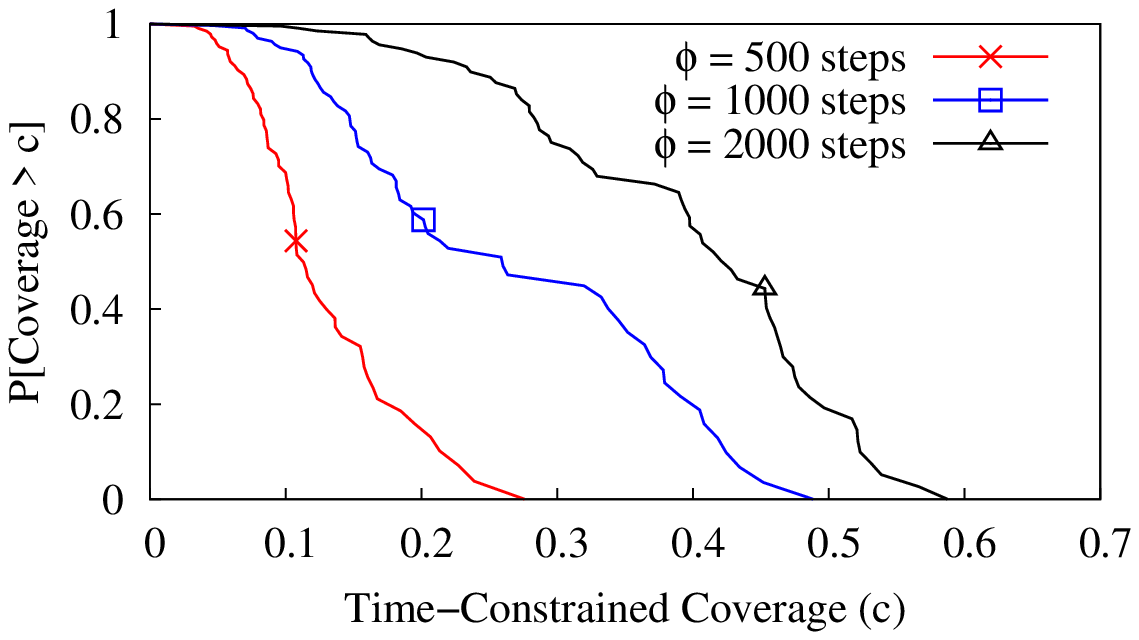}
   \label{time-constrained-distribution-mosar}
 }
\subfigure[Time-constrained coverage distribution in randomized TVG. According to the results, for a diffusion of $\Phi = 25$ time steps, the median time-constrained coverage is 3.1\% of the network. Similarly, for $\Phi = 100$ time steps, the median fraction of network reached is 43.7\%.]{
   \includegraphics[scale=0.7]{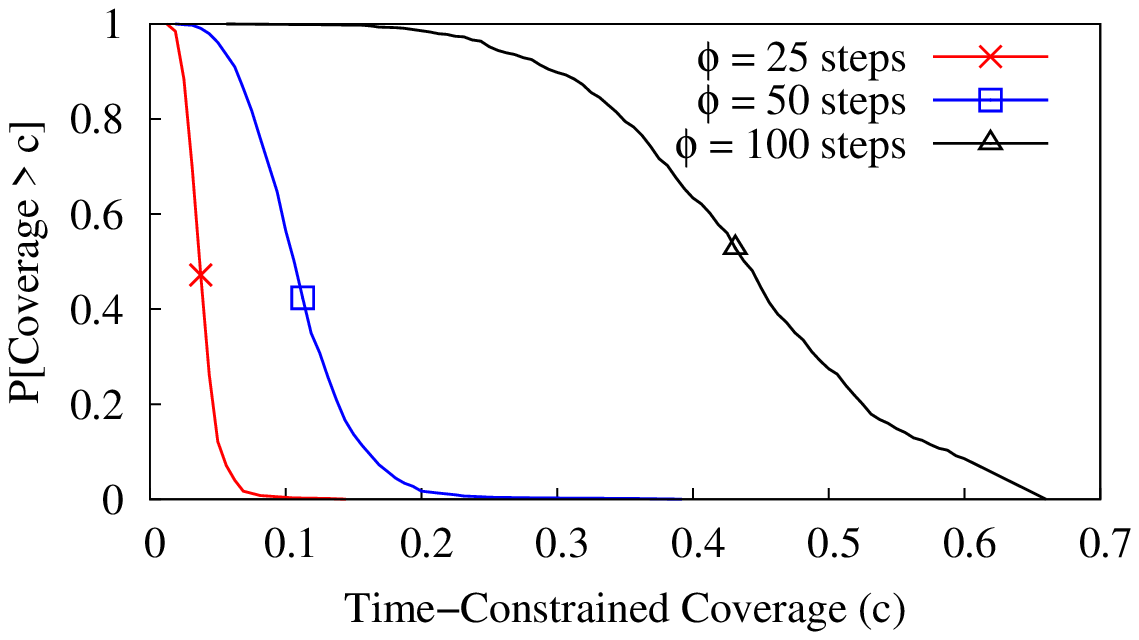}
   \label{time-constrained-distribution-Random}
 }

\caption{Time-constrained coverage distributions in MOSAR and randomized TVGs.}
\label{time-constrained-distribution}
\end{figure}

The time-constrained coverage is also able to select few time instants that present distinct diffusion performance.
Figure~\ref{time-constrained-distribution} presents the complementary CDF of the time-constrained coverage 
of the MOSAR and randomized TVGs for different values of $\Phi$ time steps. Likewise in cover time of the MOSAR TVG, the first 33000 time instants are compared on each case, while the first 660 time instants are compared on each case of the randomized TVG.

In Figure~\ref{time-constrained-distribution}, each value in a curve represents the fraction of the studied time instants that achieved at least the corresponding time-constrained coverage~$c$ of the TVG in $\Phi$ time steps.
For example, in Figure~\ref{time-constrained-distribution-mosar}, with a limit of $\Phi = 2000$ time steps diffusion starting at 80\% of the time instants reach at least 28\% of the nodes in the MOSAR TVG, noting that the maximum reachable coverage is 58\%. 

\begin{figure}[t]
\centering

\subfigure[Time-constrained coverage in the MOSAR TVG. The median time-constrained coverage for a $\Phi = 2000$ time steps diffusion is 56\%, if the diffusion starts at the top 10 time instants; whereas for 10 randomly chosen time instants, time-constrained coverage is 36\%.]{
   \includegraphics[scale=0.7]{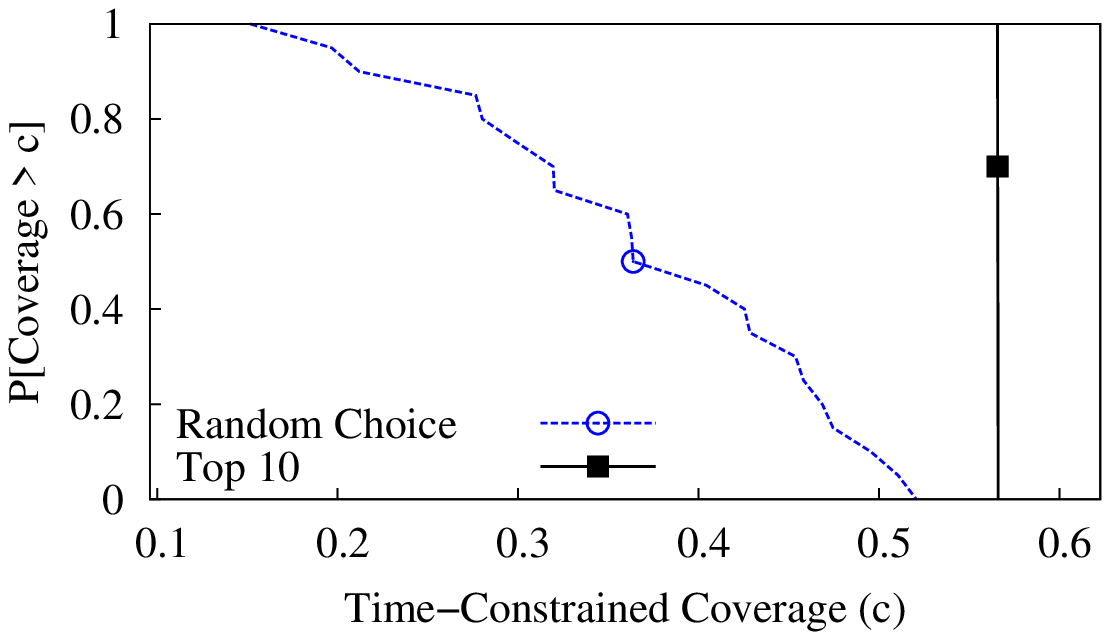}
   \label{constrained-time-top10-mosar}
 }

\subfigure[Time-constrained coverage in the randomized TVG. The median time-constrained coverage for a $\Phi = 100$ time steps diffusion is 60\%, if the diffusion starts at the top 10 time instants; whereas for 10 randomly chosen time instants, time-constrained coverage is 52\%.]{
   \includegraphics[scale=0.7]{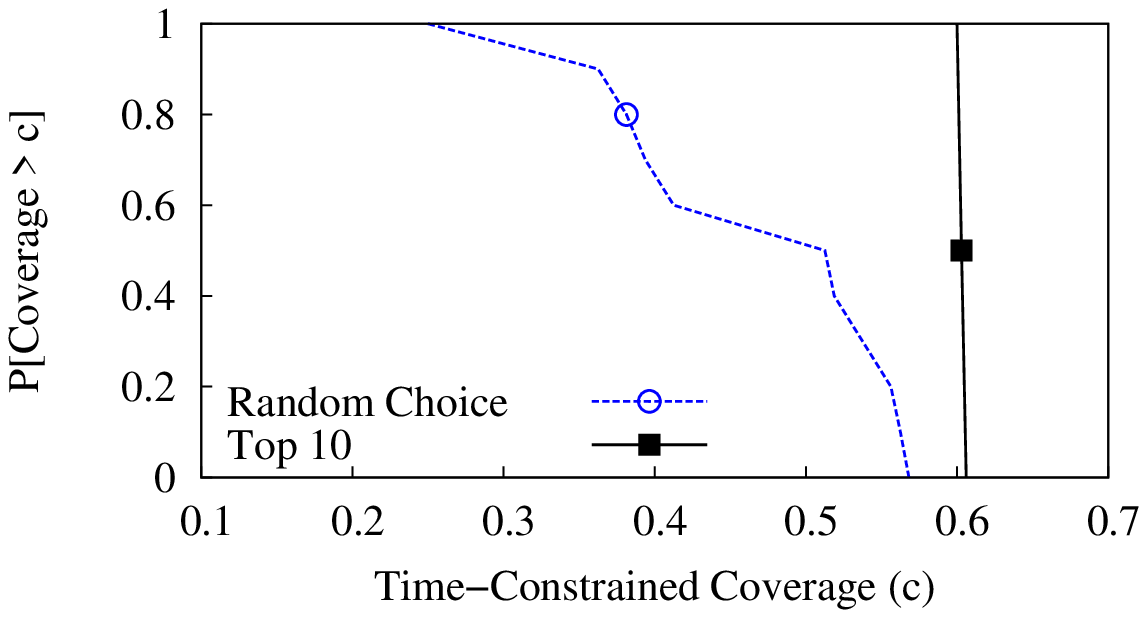}
   \label{constrained-time-top10-random}
 }

\caption{Time-constrained coverage in MOSAR and randomized TVGs: top 10 vs Random Choice.}
\label{TVG-coverage-top10xRandom}
\end{figure}

As expected, the larger is the maximum number $\Phi$ of allowed time steps, the larger is the overall network coverage. Importantly, however, in all cases a very limited number of time instants --- actually, the most central ones --- provide superior efficiency in covering more nodes in both the MOSAR and the randomized TVGs with a constrained number of time steps when diffusion starts at these most central time instants. In Figure~\ref{TVG-coverage-top10xRandom}, for both the MOSAR and the randomized TVGs, we compare the performance of the top 10 most central time instants according to the time-constrained coverage metric against 10 randomly chosen time instants to start spreading a diffusion. Indeed, in Figure~\ref{constrained-time-top10-mosar}, starting a diffusion on any of the top 10 most central time instants would reach at least 56\% of network coverage, whereas in the remaining randomly chosen time instants, a diffusion process would reach no more than 33\% of the MOSAR TVG nodes, considering a limit of $\Phi=2000$ time steps.

\subsection{Further Discussion}

Overall, the observed time centrality results, for all considered scenarios, do not follow a normal or any other narrow distribution. 
Indeed, this has been evidenced in Figures~\ref{cover-time-top10xRandomChoices} and~\ref{TVG-coverage-top10xRandom},
where we clearly note that random choices of time instants present results for the considered metrics that are significantly worse than the top 10 ranked time instants. This kind of behavior typically indicates heavy-tailed distributions, thus highlighting the relevance of time centrality in distinguishing the relative importance of different time instants for diffusion purposes.

In the experiments, both the MOSAR and the randomized TVGs have the same number of nodes and a similar node density. Despite that, we have observed that the time centrality results differ in magnitude when we compare both the results for the MOSAR and the randomized TVGs, although they present a similar general behavior. 
This difference in the magnitude of results is a consequence of the contact opportunities nodes have in the considered TVGs. In fact, in the randomized TVG, more than 99\% of edges shift from connected to disconnected (and vice-versa), thus providing uniform contact opportunities for virtually all nodes, allowing faster as well broader diffusion. In contrast, the dynamic in-person contact network represented by the MOSAR TVG presents only 56\% of edges shifting their current states, thus significantly limiting the contact opportunities nodes have to diffuse information. As a consequence, diffusion becomes much slower and constrained in the MOSAR TVG as compared with the randomized TVG.

%\corrigido{Furthermore, our results indicates that highly dense networks do not present a well defined time centrality ranking.
%In this case, ranking metric difference between nodes do not allow us to clearly identify important time instants.
%In sum, we believe that our time centrality concept can be better applied on sparse and low dense networks, which correspond to a large variety of real networks.}

%limitation
Finally, we highlight that, in this paper, the intended focus was mainly on pointing out the relevance of the time centrality concept. We thus expect this paper to motivate further research on how to detect central time instants using past or present information. This opens perspectives to investigate suitable metrics for such an evaluation as well as detection systems of central time instants based on those metrics. These natural next steps in the way the paper points out are our intended future work.

\section{Summary and Outlook}
\label{conclusion}

%\begin{enumerate}
	%\item Recap what you did. In about one paragraph recap what your research question was and how you tackled it.
	%\item Highlight the big accomplishments. Spend another paragraph explaining the highlights of your results. These are the main results you want the reader to remember after they put down the paper, so ignore any small details.
%\item Conclude. Finally, finish off with a sentence or two that wraps up your paper. I find this can often be the hardest part to write. You want the paper to feel finished after they read these. One way to do this, is to try and tie your research to the “real world.” Can you somehow relate how your research is important outside of academia? Or, if your results leave you with a big question, finish with that. Put it out there for the reader to think about to.
%\item  Before you conclude, if you don’t have a future work section, put in a paragraph detailing the questions you think arise from the work and where you think researchers need to be looking next.	
%\end{enumerate}

In this work, we introduce the notion of \emph{time centrality} in dynamic complex networks. Time centrality assesses the relative importance of a given time instant within a time-varying graph~(TVG). In this context, we present two time centrality metrics focused on diffusion processes and evaluate them using a real-world dataset representing a dynamic in-person contact network.
Our results show that starting a diffusion at the most central time instants, according to our centrality metrics, can make diffusion processes in TVGs faster and more efficient. 
%\art{Revisar paragrafo a partir daqui} For example, to address 10\% of network, top 10 Cover Time instants can make diffusion process in less than 210 (t) while in remaining instants, diffusion process would take up to 3.07 times longer. For Time-Constrained Coverage, starting diffusion on top 10  time instants would reach the at least 56\% of network coverage, while in more than remaining instants, diffusion process would reach no more than 33\% of network, considering 2000 steps of algorithm.

Considering the notion of time centrality in dynamic complex networks opens several perspectives for further research. 
As future work, we plan to further analyze the particular case of considering time centrality conditioned upon the seed node that holds the information to be diffused. Furthermore, we also intend to evaluate the trade-off on the diffusion performance in TVGs between waiting for the most central time instant to spread information or starting the diffusion before the best moment with better performance expectations considering the total time of both retaining and spreading of information. 

Moreover, in our future work we plan to develop prediction models based on time centrality for complex systems that can be represented by TVGs. In that sense, strategies for the possible early identification of central time instants might be based on discovering evidence in the recent past or the present moment of the TVG evolution indicating that a given time instant is relatively better (i.e. more central) with respect to a set of others. 

%Moreover, we also intend to conceive as future work prediction models based on time centrality for complex systems that can be represented by TVGs. In that sense, creating strategies based on monitoring and analysis of the TVG dynamics is also needed. Such strategies are key to discover evidence in the recent past or the present moment of the TVG evolution indicating that a given time instant is relatively better (i.e. central) with respect to a set of others. 
%The possible early identification of central time instants thus requires a prior knowledge of the recent network state and its evolution over time. This is, however, not a limitation since many TVGs model systems with notoriously cyclic behavior, such as transportation networks.

% Artur: achei um tanto reducndante 
%We have intuited that these evidences may be found in patterns of connection establishment or in appearance or movement of nodes over time.Based on important time instants properties in the past, since these properties occur again, the concerned prediction model would be able to indicate a potentially significant time instants.

\section*{Acknowledgments} 

This work had the support of CAPES, CNPq, FAPEMIG, and FAPERJ. Authors are grateful to Eric Fleury (ENS-Lyon/INRIA, France) for kindly providing us the MOSAR dataset we use in this work.

\bibliographystyle{abbrv}
\bibliography{ws-acs-revised}
	
\end{document}